\documentstyle[eqsecnum,floats,prd,aps]{revtex}

\begin{document}
\draft

\title{Flat spots: topological signatures of an open universe in COBE sky maps}
\author{Janna J. Levin${}^1$, John D. Barrow${}^{1,2}$, 
Emory F. Bunn${}^{1,3}$, and Joseph Silk${}^1$}
\address{$^{1}$Center for Particle Astrophysics,
UC Berkeley, Berkeley, CA 94720-7304}
\address{$^{2}$Astronomy Centre, University of Sussex, 
Brighton BN1 9QH, U.K.}
\address{$^{3}$
Physics and Astronomy Department,
Bates College,
Lewiston, ME  04240}
\address{({ submitted January 16, 1997 })}\par
\twocolumn[
\maketitle
\widetext
\begin{abstract}
We investigate the behaviour of light rays in an open
universe with a partly periodic horn topology. The geodesics
can be solved exactly and the periodic topology creates
characteristic new effects in temperature maps of the microwave background.
Large flat regions appear in COBE sky maps even when multiple
images of astronomical sources are unobservable. 
\end{abstract}
] \narrowtext
\begin{picture}(0,0)
\put(410,190){{ CfPA-97-TH-01}}
\end{picture} \vspace*{-0.15 in}
\setcounter{section}{1}

Despite the appeal of a nearly flat cosmology, the universe may have a
significant negative curvature. Detailed studies have begun to compare the
predicted large scale structure of open universes with observations \cite
{{steb},{lyth},{gorski},{whitesilk}}. We can also learn about the global
topology of space from such a comparison. The topology is not directly
determined by the matter content of the Universe: indeed, it is not fixed by
the Einstein equations at all. Yet, astronomical observations are influenced
by the Universes's topology and the cosmic microwave background radiation
(CBR) provides a uniquely sensitive probe. We present some new observable
topological effects on the CBR that occur in open universes with
non-standard topologies. These effects can be observed in the absence of
ghost images.

The topologies of a flat universe are very limited and have been studied by
many authors \cite{flat}. Remarkably, the 3-torus was first considered by
Friedmann himself in 1923 \cite{fried} in order to counter Einstein's claim
that universes with finite volume have positive curvature.
Negatively curved spaces can support a richer spectrum of possible
topologies. Negatively curved spaces create hyperbolic geodesic
motions as neighboring geodesics deviate from each other exponentially
rapidly: there is sensitivity to initial conditions.
When the space is compactified these trajectories
mix and fold, creating a structurally stable chaotic motion \cite{balazs}.
Therefore there are no analytic solutions for the motions of the CBR
photons. However, 
potentially generic features of
compact
universes may appear in the statistics of ghost images in the CBR \cite{gs}. 
By contrast, if some dimensions are compact while others remain unbounded,
the motion may be integrable with exact solutions that display the effects
of the compact directions on local geodesics. We consider non-compact
topologies of this integrable variety, motivated by the simplicity of
regular motion and the strong observable features that they create. We find
exact solutions for the geodesic motions on an open space with the topology
of a horn, introduced by Sokolov and Starobinsky in 
\cite{sos}.

Limits on any universal topology's periodicity scale are often set by
searching for periodicity in observations of large structures  \cite{flat}.
Stevens, Scott, and Silk \cite{sss} pointed out that, in a flat $3$-torus
universe, a much stronger lower bound could be set
using the CBR power spectrum. The compact space imposes a mode cutoff, which
is not observed, and constrains the minimum periodicity scale to exceed the
particle horizon. In an open horn topology there is also a cutoff in 
two directions but none along the horn.
The power
spectrum is suppressed but not so dramatically as in the flat $3$-torus
universe. More prominent features result from suppression of the power
spectrum along the narrowest part of the horn. This occurs even if the
topology scale is so large that no periodicity of images is visible. We will
show that this suppression leads to flat spots in CBR sky maps. However,
chaotic geodesic mixing has been avoided at the cost of destroying both
homogeneity and isotropy. Only the largest modes are affected since only
these will probe the topology scale. Consequently, a contribution to the
lowest multipoles of the angular power spectrum is prominent in the maps.

Consider an open Friedmann metric
\begin{equation}
ds^2=a^2\left[ -d\eta ^2+dr^2+\sinh ^2r\left( d\theta ^2+\sin ^2\theta d\phi
^2\right) \right]  \\
.
\end{equation}
Make the coordinate transformation \cite{sos}, 
$e^{-z}=\cosh {r}-\sinh {r}\cos {\theta },e^{-z}x=\sin {\theta }\cos {\phi }
\sinh {r},e^{-z}y=\sin {\theta }\sin {\phi }\sinh {r}$, so the metric
becomes 
\begin{equation}
a^{-2}ds^2=-d\eta ^2+dz^2+e^{-2z}(dx^2+dy^2).  \label{one}
\end{equation}
The topology is induced by identifying points periodically along $x$ and $y$
by $(x,y)\equiv (x+b,\ y+h)$, where $b$ and $h$ are constants, to create a $2
$-torus. This torus is stretched or shrunk by the factor $e^{-2z}$ along the 
$z$-axis to create a toroidal horn. 
The comoving proper area of the torus is 
\begin{equation}
\ \int_0^adx\int_0^bdy\ e^{-2z}=e^{-2z}bh\ \ ,
\end{equation}
and depends on location along the $z$-axis. The global topology clearly
introduces global inhomogeneity as well as global anisotropy.

Primordial photons scatter off perturbations in the background curvature. In
an open universe the Sachs-Wolfe effect, which produces the CBR temperature
perturbation $\delta T/T$, receives a contribution from the surface of last
scattering and from a cumulative integral over the gravitational potential
taken along a geodesic: 
\begin{equation}
{\frac{\delta T}T}={\frac 13}\Phi (\eta _o\hat n)+2\int_o^{\eta _o}d\eta
\Phi (\eta \hat n)^{^{\prime }}\ ,
\end{equation}
where $^{\prime}\equiv d/d\eta $. The potential $\Phi (\eta _o\hat n)$ is to
be evaluated somewhere in the volume of last scattering. We need to locate
in that volume the sphere of radius $\eta _o$ we see in direction 
$\hat n(\theta ,\phi )$. The most general geodesics can be located \cite{us}
but we
only need the radial geodesics in spherical coordinates. The periodicity of
the manifold will be accounted for by imposing periodic boundary conditions
on the eigenfunctions of the potential. We receive light rays along the lines 
\begin{eqnarray}
e^{-z} &=&\cosh (\eta _o-\eta )-\sinh (\eta _o-\eta )\cos \theta   \nonumber
\\
e^{-z}x &=&\sin \theta \cos \phi \sinh (\eta _o-\eta )  \nonumber \\
e^{-z}y &=&\sin \theta \sin \phi \sinh (\eta _o-\eta ).  \label{ge}
\end{eqnarray}
Here, $\theta $ and $\phi $ are angles defining the direction $\hat n(\theta
,\phi )$ in which the photon is observed. Parametrically, the photon path is 
\begin{equation}
e^{2z}=W_{\rm i}^{-2}-(x-\hat x_{{\rm i}})^2-(y-\hat y_{{\rm i}
})^2\ \ .
\end{equation}
The constants $W_{{\rm i}},x_{{\rm i}},y_{{\rm i}}$ can be related to $\hat n
(\theta ,\phi )$ by $W_{{\rm i}}^2=\sin ^2\theta $, $\hat x_{{\rm i}}=\cos
\phi \cot \theta $, and $\hat y_{{\rm i}}=\sin \phi \cot \theta $. If we
identify the topology by ($x,y)\equiv (x+b,$ $y+h)$, the photons will
spiral around the flat torus in $(x,y)$ while the entire torus is
hyperbolically stretched as light moves along $z$. The most general photon
motion will be a looping spiral. These trajectories have surprising
properties. Photons never travel to $z=+\infty $ unless they start there (or
move along the $z$-axis). Any photon travelling with increasing $z$
eventually hits a maximum and then wraps back. As $W_{{\rm i}}\rightarrow 0$
, the geodesics tend to straighten out. If $W_{{\rm i}}^2=0$ there is no
motion along $(x,y)$ and the trajectories are the lines $z=\pm \eta +z_{{\rm 
i}}$. 

The gauge-invariant gravitational potential perturbation, $\Phi ,$ can be
expanded in terms of eigenfunctions 
\begin{equation}
{\Phi }=\int dk\ \sum_w\sum _n\Phi _{kwn}(\eta )\psi _{kwn}(x,y,z),
\end{equation}
where the $\psi _{kwn}$ are spatial eigenfunctions and the $\Phi _{kwn}(\eta
)$ are the time-dependent amplitudes. The cosmological perturbations satisfy
the scalar wave equation $\Phi ^{\prime \prime }+2{\cal H}\Phi -\Delta \Phi ,
$ with ${\cal H}\equiv a^{\prime }/a$. Following \cite{lyth}, we look for
separable solutions with $\Phi _{kwn}=\hat \Phi _{kwn}F(\eta ).$ The $\hat 
\Phi _{kwn}$ is a primordial time-independent perturbation predicted from,
say, inflation; $F(\eta )$ describes the time evolution with
\begin{equation}
F(\eta )=5{\frac{\sinh ^2\eta -3\eta \sinh \eta +4\cosh \eta -4}{(\cosh \eta
-1)^3}}\ \ .
\end{equation}
The spatial eigenmodes can be found in the $(x,y,z)$ coordinate system \cite
{sos}.  If
\begin{equation}
{\frac{\delta T}T}(\hat n)=\int_1^\infty dk\ \sum_{wn}\Phi _{kwn}(\eta _o
\hat n)N_{kwn}L_{kwn}  \label{it}
\end{equation}
with the normalization 
\begin{equation}
N_{kwn}=\left( {\frac{2k\sinh (\pi k)}{\pi ^2}}{\frac{(2-\delta
_{w0})(2-\delta _{n0})}{bh}}\right) ^{1/2}
\end{equation}
then the spatial eigenfunctions are
\begin{eqnarray}
L_{kwn}=\left[ {\frac 13}\right.  &+&\left. 2\int_{\eta _{{\rm i}}}^{\eta
_o}d\eta F^{\prime }(\eta )\right] e^zK_{ik}(Qe^z)\times   \nonumber \\
&&\ \ \ \pmatrix{ \sin\left({2\pi w\over b}x\right )\cr \cos\left({2\pi
w\over b}x\right )\cr}\pmatrix{ \sin\left({2\pi n\over h}y\right )\cr
\cos\left({2\pi n\over h}y\right )\cr}\ \ 
\end{eqnarray}
where $K_{ik}$ is a modified Bessel function with imaginary index and the
entire function is included here in the integration over $\eta $. The
quantity $Q$ in the argument of the Bessel function is related to the
discrete eigenmodes through $Q^2=4\pi ^2(w^2/b^2+n^2/h^2)$. The integral is
taken from the time of last scattering, $\eta _{{\rm i}}$, until today, 
$\eta _o$. We have not included the supercurvature modes in the integral in
eqn (\ref{it}) since they were not found to contribute significantly.

Using the convention of \cite{lyth}, the primordial fluctuation amplitude is
assigned an independent Gaussian probability distribution 
\begin{equation}
\left\langle \hat \Phi _{kwn}^{*}\hat \Phi _{kwn}\right\rangle ={\frac{2\pi
^2}{k|k^2-1|}}{\cal P}_\Phi (k)\delta (k-k^{\prime })\delta _{ww^{\prime
}}\delta _{nn^{\prime }}\ \ .  \nonumber  \label{assume}
\end{equation}
The prediction of de Sitter inflation is that ${\cal P}_\Phi $
is a number. We assume all modes are statistically independent and
equi-probable. In a homogeneous universe, the amplitude of perturbations
will be the same everywhere, on average. By contrast, in our model
the amplitude decays inside the small regions of the horn \cite{sos}.
Although the amplitude of each mode is assigned a value assuming Gaussian
statistics, the global anisotropy may not be Gaussian.

Curvature and topology effect wavelengths long compared to the diameter of
the torus in a given direction. The long-wavelength limit is $Qe^z\gg k,$
where $K_{ik}(Qe^z)\simeq \sqrt{\frac \pi {2Q}}\exp \left( -z/2-Qe^z\right) $. 
The sum in $\delta T/T$ is quickly suppressed as $w$ and $n$ are
increased. The largest contribution to the sum comes from the smallest value
of $Q$. Setting $b=h$ gives a minimum $Q$ value of $2\pi /b$ for $(w=1,n=0)$
and $(w=0,n=1)$. Consequently, we have 
\begin{eqnarray}
\left\langle \left| {\delta T/T}(\hat n)\right| ^2\right\rangle  &=&{\frac{
2F^2}{9b}}\exp \left( z-(4\pi /b)e^z\right) \pmatrix{ \sin\left({2\pi \over
b}x\right )\cr \cos\left({2\pi \over b}x\right )\cr}^2  \nonumber \\
&\times &\int_0^{Qe^z}dk{\frac{{\cal P}_\Phi (k)\sinh (\pi k)}{|k^2-1|}}
+(x\leftrightarrow y)+\ \ ...  \nonumber
\end{eqnarray}
For simplicity we have omitted the integrated Sachs-Wolfe term. Along the
axis in which the space broadens we have $Qe^{-\Delta \eta }\simeq 0.1$
when $\Omega_0=0.3,b=h=5$. For
our approximation to be valid we need $k\ll 0.1$ and only the supercurvature
modes are suppressed. The full set of subcurvature modes are unsuppressed
and so should produce a normal spectrum in that direction. Along the neck of
the horn, however, $Qe^{\Delta \eta }\simeq 13.8,$ and so many modes will be
affected. From the above expansion the spectrum  
along the narrow axis of the horn, $\hat k$, is suppressed by a factor 
\begin{equation}
\propto \exp \left[ \Delta \eta -(4\pi /b)\exp \Delta \eta \right] \ \ .
\label{try}
\end{equation}
For a topology scale of $b=h=5$ and a cosmological density parameter of $
\Omega _0=0.3$ today ($\Delta \eta \simeq 2.398$), the suppression factor is
about $10^{-11}$. In the narrow part of the horn, the sky temperature map is
therefore very much flatter. This suppression is partially camouflaged. Much
of the CBR we see originated in the wide part of the horn, travelled down
the neck, and wrapped back so that it appears to come from the narrow part
of the horn. When we look along the squeezed direction (small $\theta $),
the light  we see only truly originated in the neck for $\cos \theta \mathrel
{\hbox to 0pt{\lower 3pt\hbox{$\mathchar"218$}\hss}\raise 2.0pt
\hbox{$\mathchar"13E$}}(\cosh \Delta \eta -1)/\sinh \Delta \eta $. For $
\Omega _0=0.3$ this corresponds to $\cos \theta \mathrel{
\hbox to
0pt{\lower 3pt\hbox{$\mathchar"218$}\hss}\raise 2.0pt\hbox{$\mathchar"13E$}}
0.83$. The rest of the structure seen in that direction is reflected from
the wide part of the horn. Like in a hall of mirrors, nothing is where it
appears. 

On concentric rings centered on the $z$-axis, a range of small $k$ modes
will be suppressed when $Qe^z\gg 1$. These rings subtend a half angle in the
sky with 
\begin{equation}
\cos \theta >{\frac{\cosh \Delta \eta -Q}{\sinh \Delta \eta }}\ \ ,
\end{equation}
where we have used the geodesic equations (\ref{ge}). The unsuppressed modes
give structure on a 
scale $\lambda (\theta )\mathrel{\hbox to
0pt{\lower 3pt\hbox{$\mathchar"218$}\hss}\raise 2.0pt\hbox{$\mathchar"13C$}}
(\cosh \Delta \eta -\sinh \Delta \eta \cos \theta )/Q$. Therefore, we expect
to see rings of smaller and smaller lumps as one moves toward the axis. The
smaller the topology scale, the smaller the lumps, and the broader the
affected patch of sky.

In Fig. 1 we give a predicted COBE map for $b=h=1$ and 
$\Omega _0=0.3$. To render the generation of the maps numerically tractable
we only sample a finite number of modes. This leads to a completely flat
spot where realistically we would expect to see very small structures. 
When a maximum $k$ is imposed, the flat spot
should subtend an angle $\cos \theta \mathrel{
\hbox to 0pt{\lower
3pt\hbox{$\mathchar"218$}\hss}\raise 2.0pt\hbox{$\mathchar"13E$}}(\cosh
\Delta \eta -Q/k_{{\rm max}})/\sinh \Delta \eta $.  For the maps shown here,
modes up to $k=10$ are included so that the completely flat spot is
predicted to subtend a half-angle of roughly $26^{\circ}$ for $b=h=1$ and 
$\Omega_0=0.3$. This agrees with the size of the feature in the
top panel of Fig. 1. The small structure on concentric rings discussed above
can also be seen most prominently surrounding the spot but the effect
extends across half the sky. The lower two panels of Fig. 1 show
predicted COBE skies when the topology scale and the cosmic density are varied.
The lowest of these shows the long-wavelength 
temperature variations along the $z$-axis superimposed 
on the small fluctuations
permitted along the torus.  In the map shown, the torus is oblong
as the topology scales are unequal.

The dilution of the power spectrum we have found can be related to an
analogous flat space result. In a flat toroidal
universe, discrete modes with infinitely small $w$ and $n$ cannot fit inside
the finite torus: there is a 
maximum wavelength $\lambda _{max}=b$ \cite{sss}. Similarly, for the open
horn-shaped space, there is a long-wavelength mode cutoff in the $x$ and $y$
directions and a suppression of the power spectrum at small $k$ as seen in
eqn (\ref{try}).

For comparison, the angular power spectrum $C_\ell$ in an anisotropic
cosmology is
\begin{eqnarray}
(2\ell +1)C_\ell =\sum_m &&\int d\Omega ^{\prime
}Y_{\ell -m}(\hat n^{\prime })\int d\Omega Y_{\ell m}(\hat n)  \nonumber \\
&&\ \ \times  \left\langle {\delta T/T}(\hat n^{\prime })
{\delta T/T}(\hat n)\right\rangle \ \ .
\end{eqnarray}
In the absence of rotational symmetry, the sum over $m$ does not collapse
simply.  The power spectrum can be estimated by expanding an individual
simulated sky map in spherical harmonics and computing the mean-square
amplitude of the coefficients at a given $\ell$.  The results of such a 
calculation are shown in Fig. 2.  As can be seen in that figure, the
flat region gives a sharp dip in the power spectrum at $\ell$'s
that correspond roughly to the size of the flat spot on the sky.

The spectra in Fig. 2 have reductions in power at low $\ell$,
when compared to a flat Harrison-Zel'dovich spectrum
($C_\ell\propto 1/\ell(\ell+1)$).  The dips in general 
mean these models fit the COBE DMR data quite poorly.
For example, both
of the $\Omega_0=0.3$ models have COBE likelihoods roughly 20 times less
than that of a flat spectrum.  (These likelihoods were computed using
the Karhunen-Lo\`eve method described in \cite{bunnwhite}.)
Since our models are not described by isotropic
Gaussian statistics, we cannot draw precise conclusions from these results,
but it seems safe to say that these models provide a poor fit to the DMR
data.

Topological effects on the geodesics are extremely noticeable even
though most trajectories do not wind around the universe more than once. 
For $\phi =0$, the maximal number of windings is executed by trajectories in
the direction $\cos \theta =\tanh (\Delta \eta )$ for which $m_x=[\Delta
x/b]=[\sinh (\Delta \eta )/b]=1$ for $b=5$ and $\Omega _0=0.3,$ (
square brackets denote the integer part). Only paths in this
vicinity begin to wind more than once. Furthermore, a multiplicity of
windings does not guarantee periodicity in this topology. True periodicity
requires $x=x+m_xb$ {\it and simultaneously} that $y=y+m_yh,$ where $m_x$
and $m_y$ are the integer winding numbers defined above. The location in the
sky, $\hat n(\phi )$, of any periodic images can be ascertained by imposing
these boundary conditions on eqns (\ref{ge}) for fixed $\theta $. The
resulting conditions are not generally easy to satisfy. 

Other effects can destroy perfect periodicity, which indicates the weakness
of periodicity constraints. For example, as a consequence of the integrated
Sachs-Wolfe effect, photons travelling along different paths will suffer
different distortions, even if they originated in the same place. Likewise,
any net lensing of the photon background will smear out periodicity.

The decay of the perturbation amplitude can have further consequences
The distribution of galaxies should also be
affected \cite{sos}. Only small structures could form in the narrow part of
the horn. 
The global topology
may then create
local inhomogeneity 
and distinctive patterns in the large-scale distribution of luminous and dark
matter. 
%This in turn could warp the topology. 

We have found that some topologies create strong effects on CBR sky maps in
open universes. The periodicity scale for the horn topology we considered
must exceed the size of the observable universe in order to hide the flat
spots otherwise predicted in these maps. The horn
configuration is a particular case, chosen to avoid chaotic non-integrable
geodesics, yet it is a good model of some characteristic features of complex
identified topologies. In particular, the suppression of the power spectrum
or features analogous to flat spots on the microwave sky 
will likely be more incisive in identifying topology
than ghost images \cite{comment}.
We have shown these attributes provide a sensitive probe of periodicity 
in the global topology of the Universe.

We thank J.R. Bond, P. Ferreira and A. Jaffe for valuable discussions. JDB
is supported by the PPARC and acknowledges support from the Center for
Particle Astrophysics, Berkeley.  This research has been supported in 
part by a grant from NASA.

\vfill\eject
\newpage

\
\begin{figure}[h]
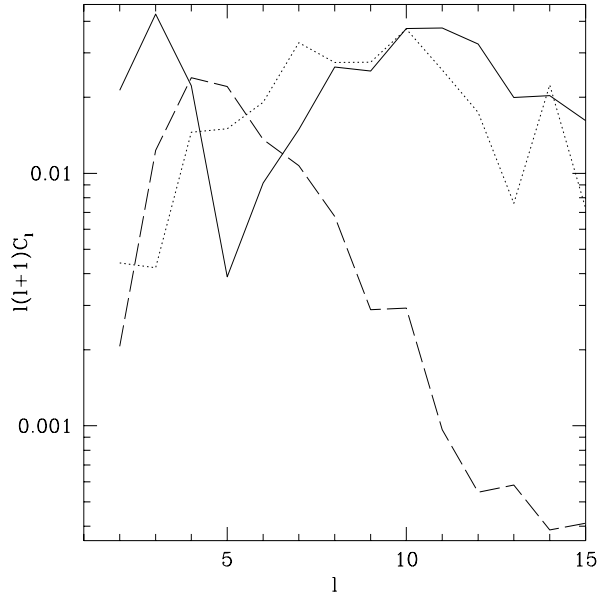

\vspace{100mm}
\includegraphics{top1_k10n25nd15_mimas.ps}
\includegraphics{top1_k10n25nd15_.6_mimas.ps}
\includegraphics{top.3b1_k10n50nd20_coma.ps}
\vspace{20mm}
\caption{Three maps of the predicted COBE sky in an open universe
wrapped into a horn.  The topology scales and cosmic density 
are $b=h=1$ and $\Omega_0=0.3$ in 
the top panel.  The middle panel 
has $b=h=1$ with $\Omega_0=0.6$.
Asymmetric topology scales $b=0.3,h=1$ in an $\Omega_0=0.3$
horn lead to an oblong flat spot as seen in the bottom map.
When one or both of the scales is small,
the temperature variation along $z$ is superimposed on
the small fluctuations along $x$ and $y$.}
\vspace{8mm}
\end{figure}

\
\begin{figure}[h]
\vspace{40mm}
\includegraphics{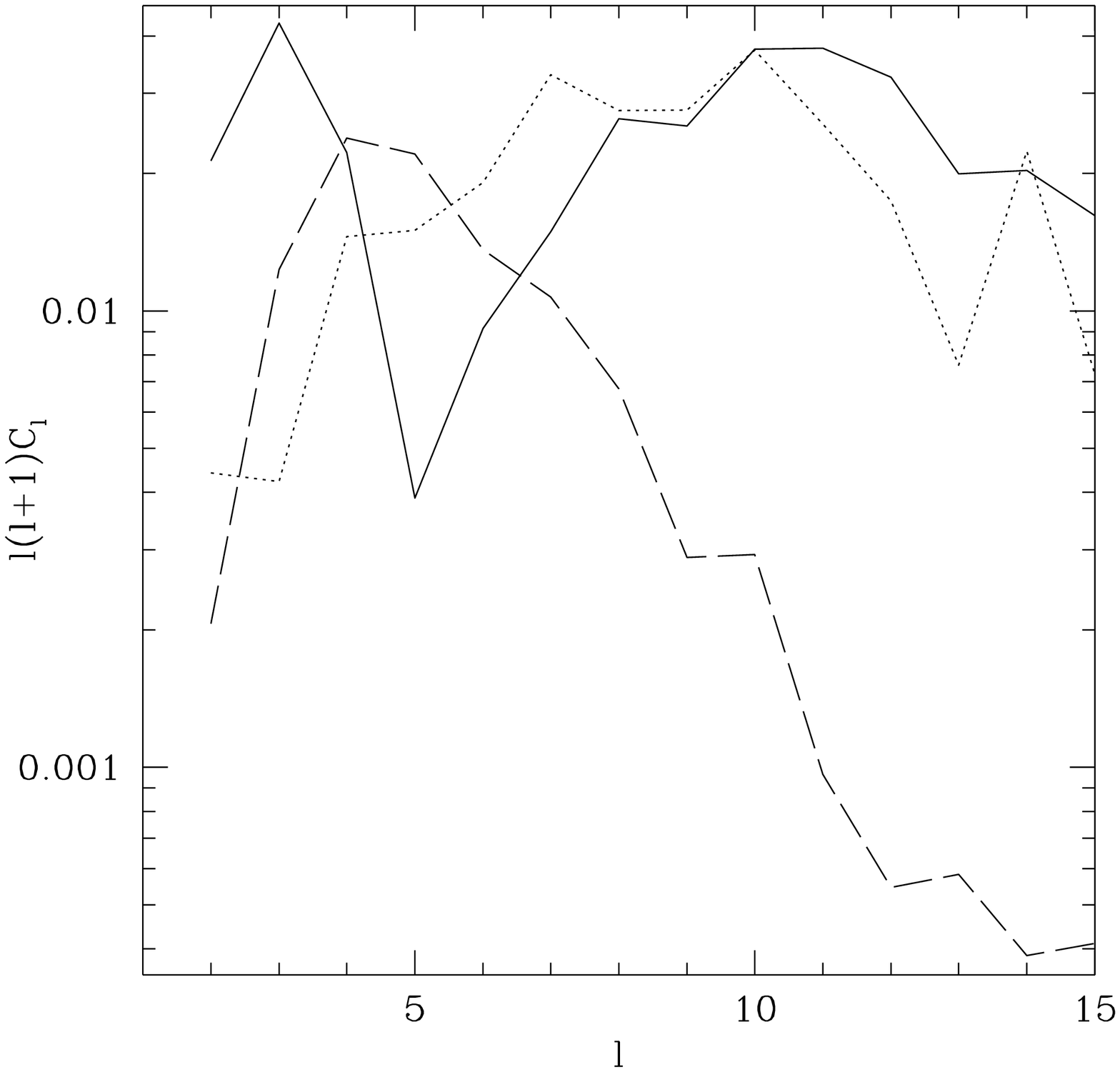}
\vspace{8mm}
\caption{Angular power spectra.  The solid line corresponds
to $b=h=1, \Omega_0=0.3$.  
The dotted line is the asymmetric case of $b=0.3,h=1$ and a
density of $\Omega_0=0.3$
The dashed line depicts $b=h=1, \Omega_0=0.6$.}
\end{figure}


\begin{references}
\bibitem{comment} The proliferation of
periodic orbits in the ergodic case of a compact pseudosphere might
have the opposite effect of suppressing the power spectrum at small $k$.
Still, if the complete basis of subcurvature modes ($k\ge 1$)
are sufficient to describe a perturbation on a compact, open space
then the exponential growth of $k\rightarrow 0$ modes may not 
be relevant.  This is a subtle issue worthy of further attention.
\end{references}
\end{document}